\documentclass[lettersize,journal]{IEEEtran}
\usepackage{amsmath,amsfonts,amssymb}
\usepackage{algorithmic}
\usepackage{booktabs}
\usepackage{color}
\usepackage{ulem}
\usepackage{algorithm}
\usepackage{array}
\usepackage[caption=false,font=normalsize,labelfont=sf,textfont=sf]{subfig}
\usepackage{textcomp}
\usepackage{stfloats}
\usepackage{url}
\usepackage{multirow}
\usepackage{verbatim}
\usepackage{graphicx}
\usepackage{cite}
\usepackage{CJKutf8}
\usepackage{soul}
\usepackage{pifont}

\DeclareMathOperator*{\argmax}{argmax}
\begin{document}
\begin{CJK}{UTF8}{gbsn}
\title{Aligning Speech to Languages to Enhance Code-switching Speech Recognition}

\author{Hexin Liu, Xiangyu Zhang, Haoyang Zhang, Leibny Paola Garcia-Perera, Andy~W.~H.~Khong, Eng Siong Chng, Shinji Watanabe
\thanks{This work was supported in part by the Project WP6 within the Delta-NTU Corporate Lab, in part by A*STAR through its IAF-ICP programme under Grant I2201E0013,
and in part by Delta Electronics Inc. The computational work for this article was fully performed on resources of the National Supercomputing Centre, Singapore. \textit{(Corresponding authors: Hexin Liu)}

Hexin Liu and Eng Siong Chng are with the College of, Nanyang Technological University, Singapore 639798 (e-mail: \{hexin.liu; aseschng\}@ntu.edu.sg). 

Xiangyu Zhang is with the School of Electrical Engineering and Telecommunications, University of New South Wales, Sydney NSW 2052, Australia (email: xiangyu.zhang2@unsw.edu.au).

Haoyang Zhang is with the School of Software and Microelectronics, Peking University, Beijing
100871, China (email: zhang.haoyoung@stu.pku.edu.cn).

Leibny Paola Garcia Perera is with the CLSP and HLT-COE, Johns Hopkins University, Baltimore, MD 21218 USA (e-mail: lgarci27@jhu.edu).

Andy~W.~H.~Khong is with the School of Electrical and Electronic Engineering and the Lee Kong Chian School of Medicine, Nanyang Technological University, Singapore 639798 (email: andykhong@ntu.edu.sg).

Shinji Watanabe is with the Language Technologies Institute, Carnegie Mellon University, Pittsburgh, PA 15213 USA (e-mail: shinjiw@ieee.org).

}}

\markboth{IEEE Transactions on Audio, Speech, and Language Processing}
{Shell \MakeLowercase{\textit{et al.}}: A Sample Article Using IEEEtran.cls for IEEE Journals}
\maketitle

\begin{abstract}
 Code-switching (CS) refers to the switching of languages within a speech signal and results in language confusion for automatic speech recognition (ASR). To address language confusion, we propose a language alignment loss (LAL) that aligns acoustic features to pseudo-language labels learned from the ASR decoder during ASR training. This approach enables frame-level language identification without the need for frame-level language annotations. To further tackle the complex token alternatives for language modeling in bilingual scenarios, we propose to employ large language models via a generative error correction method. A linguistic hint, derived from LAL outputs and decoded hypotheses, is introduced to guide the prompting and enhance the LLM-based generative error correction for CS-ASR. The proposed methods are evaluated on the SEAME dataset and data from the ASRU 2019 Mandarin-English code-switching speech recognition challenge. The incorporation of the proposed language alignment loss improves CS-ASR performance for both hybrid CTC/attention and Whisper models on both datasets, with only a negligible increase in the number of parameters. This work also highlights the efficacy of language alignment loss in balancing primary-language-dominant bilingual data during training, with an 8.6\% relative improvement on the ASRU dataset compared to the baseline model. Performance evaluation using large language models reveals the advantage of the linguistic hint by achieving 14.1\% and 5.5\% relative improvement on test sets of the ASRU and SEAME datasets, respectively.
\end{abstract}


\section{Introduction}
\IEEEPARstart{C}{ode}-Switch~(CS) refers to the switching of languages within a spontaneous multilingual recording. Intra-sentence code-switching occurs when the language changes within a single sentence, while inter-sentence code-switching involves the switching of languages at the sentence boundaries. Unlike monolingual speech, code-switched speech presents a greater challenge for automatic speech recognition~(ASR) due to language confusion and the lack of annotated data.

Although a CS-ASR system can function like monolingual ASR by combining language-specific vocabularies~\cite{survey, espnet, mamba_in_speech}, recent works address language confusion by incorporating language information. One direct approach is to optimize the ASR and language identification~(LID) or diarization~(LD) tasks jointly~\cite{zeng19_interspeech,liu23_icassp, tseng2021mandarin, liuxsa, liu2023enhancing, dhawan23}. Here, models learn language information through backpropagation from the LID or LD branch during training~\cite{liuxsa, liu22e_interspeech}, while only the ASR output is computed during inference. Beyond joint optimization, approaches based on the bi-encoder and the mixture-of-experts method build upon the Transformer architecture~\cite{transformer, speech_transformer, bi_encoder, mary20_icassp}, where the models incorporate two encoders pre-trained on monolingual data independently to capture language-specific information. Language-specific modules have also been adopted in other architectures due to their effectiveness in distinguishing languages~\cite{tt_cs, song22e_interspeech}. In contrast to language-specific encoder modules, the language-aware decoder module has been explored to reduce multilingual contextual information via a language-specific self-attention mechanism within Transformer decoders~\cite{zhang22x_interspeech}. In addition, a conditional factorization method factorizes CS-ASR into two monolingual recognition processes before integrating multiple recognized monolingual segments into a single bilingual sequence~\cite{conditionalfactorization}. As an extension, a conditionally factorized connectionist temporal classification~(CTC) module that allows for a zero-shot setting has been proposed~\cite{zeroshot_brian,ctc}. A factored language model integrating syntactic and semantic features in a code-switched language model has also been explored to enhance CS-ASR~\cite{syntactic_LM}.

Although existing works achieve reasonable CS-ASR performance, limitations in code-switching annotations and data characteristics continue to hinder further improvements. Multilingual ASR systems often benefit from utterance-level one-hot language vectors since each utterance typically contains only one language~\cite{multilingual_onehot, uml_23}. In contrast, code-switching speech signals often contain two languages, making utterance-level language labels insufficient. To address this, detecting languages at finer granularity, particularly at the frame or token level, has been suggested as a more suitable approach to CS-ASR~\cite{zeng19_interspeech, liu23_icassp, tseng2021mandarin, liu2023enhancing, dhawan23, bi_encoder, song22e_interspeech}. Despite this, most code-switching corpora do not include ground-truth frame-level language timestamps, since the human annotation process is resource-intensive and requires expertise in bilingualism. While forced alignment offers a potential solution to generate high-granularity timestamps, code-switching speech poses greater challenges for this approach compared to monolingual speech due to increased language confusion~\cite{wang19l_interspeech, rousso2024tradition, seame}. Consequently, language labels generated through forced alignment might not be desirable in terms of accuracy and computational cost. These limitations render the application of supervised language identification for CS-ASR impractical~\cite{tseng2021mandarin}. Additionally, the accent of the primary language may bias the secondary language in bilingual code-switching speech, resulting in the two languages being auditorily similar~\cite{seame, shi2020asru}. Therefore, achieving language identification in accented speech remains challenging~\cite{wanneroy1999acoustic}.

In this paper, we propose leveraging the speech-to-language alignment to improve CS-ASR performance. Central to our approach is introducing a language alignment loss (LAL), which enriches the CS-ASR model with language information without the need for additional annotations such as human labeling or forced alignment. The LAL achieves this by utilizing low-granularity yet accurate token-level language information to explicitly guide high-granularity frame-level acoustic features. The proposed LAL relies on frame-level language information and is therefore categorized as a joint optimization approach. In addition to not requiring frame-level language annotations, the proposed method also differs from auxiliary LID approaches by incorporating token-level language proportions and auditory similarity, leading to a more nuanced alignment between language and acoustic representations. Additionally, as a by-product of LAL, frame-level language predictions can be summarized as an utterance-level hint. This linguistic hint further utilizes the language information derived from speech by facilitating the incorporation of an external large-scale language model~(LLM) in ASR through a generative error correction method~\cite{touvron2023llama, hsieh-etal-2023-distilling, chen2023hyporadise}. 

The remainder of this paper is organized as follows: Section~\ref{sec:related_work} introduces the hybrid CTC/attention model, which serves as a baseline model for the proposed method. The proposed LAL and the incorporation of the code-switching hint in external language modeling are presented in Sections~\ref{sec:lal} and \ref{sec:hint}, respectively. Datasets, model configurations, and experimental setup are described in Section~\ref{sec:data_exp_model}. We present results and analysis in Section~\ref{sec:results} before highlighting the advantages and limitations of our proposed methods in Section~\ref{sec:discuss}. Finally, we conclude our work in Section~\ref{sec:conclusion}.

\section{Preliminary}
\label{sec:related_work}

\subsection{Conformer-based hybrid CTC/attention ASR model}
We employ a Conformer-based hybrid CTC/attention ASR model as the baseline. The hybrid CTC/attention model comprises an encoder, a decoder, and a CTC module, where the decoder and CTC modules share the encoder outputs~\cite{hybrid_ctc_attention_asr}. The encoder and decoder modules comprise the Conformer encoder and Transformer decoder layers~\cite{conformer, transformer, espnet}, respectively. 

Given a speech signal, we define its acoustic features as $\mathbf{X}=(\mathbf{x}_{t} \in \mathbb{R}^{F}| t=1, \ldots, T)$ and the paired token sequence as $W=(w_n \in \mathcal{V} | n=1, \ldots, N)$, where $\mathcal{V}$ is the vocabulary, $T$ and $N$ are the lengths of the feature and token sequences, respectively, and $F$ is the dimension of the acoustic feature. The encoder generates hidden outputs $\mathbf{H}=(\mathbf{h}_{t} \in \mathbb{R}^{D}| t=1, \ldots, T^{\prime})$ from $\mathbf{X}$, which are subsequently used as inputs for the decoder and CTC modules. Here, $T^{\prime}$ is the length of the hidden output sequence, where $T^{\prime}<T$ due to subsampling, and $D$ is the dimension of the hidden output. With $\mathbf{H}$, the CTC module computes the token sequences according to the Bayesian decision theory by factorizing $p_{\mathrm{ctc}}\left(W|\mathbf{X}\right)$ as~\cite{ctc}
\begin{equation}
\begin{aligned}
\setlength{\abovedisplayskip}{4pt}
\setlength{\belowdisplayskip}{4pt}
  p_{\mathrm{ctc}}\left(W|\mathbf{X}\right)&=\sum_{Z} p\left(W|Z, \mathbf{X} \right) p\left(Z|\mathbf{X}\right)\\
  &\approx \sum_{Z} p\left(W|Z\right) p\left(Z|\mathbf{X}\right),
\end{aligned}
  \label{eq:bayesian}
\end{equation}
where $Z=(z_t \in \mathcal{V} \cup \left \{ <blank> \right \} | t=1, \ldots, T^{\prime})$ is a framewise token sequence conditioned on $\mathbf{X}$. The variable
\begin{equation}
\begin{aligned}
\setlength{\abovedisplayskip}{4pt}
\setlength{\belowdisplayskip}{4pt}
  p\left(Z|\mathbf{X}\right) &= \prod_{t=1}^{T}p\left ( z_{t}|z_{1}, \ldots, z_{t-1}, \mathbf{X}\right )\approx \prod_{t=1}^{T}p\left ( z_{t}|\mathbf{X}\right )
\end{aligned}
\label{eq:ctc_acoustic}
\end{equation}
is the acoustic model of the CTC, where the probabilistic chain rule and the conditional independence assumption have been invoked. Exploiting the Bayes' rule, the probabilistic chain rule, and the conditional independence assumption, the CTC token model is given by 
\begin{equation}
\begin{aligned}
  p\left(W|Z\right) &= \frac{p\left(W\right) p\left(Z|W\right)}{p\left(Z\right)}\\
  &= \prod_{t=1}^{T}p\left ( z_{t}|z_{1}, \ldots, z_{t-1}, W\right )\frac{p\left(W\right)}{p\left(Z\right)}\\
  &\approx \prod_{t=1}^{T}p\left ( z_{t}|z_{t-1}, W\right )\frac{p\left(W\right)}{p\left(Z\right)}.
\end{aligned}
\label{eq:ctc_token}
\end{equation}
The scaling term $p\left(W\right)/p\left(Z\right)$ is often excluded, and hence~(\ref{eq:bayesian}) can be rewritten as
\begin{equation}
  p_{\mathrm{ctc}}\left(W|\mathbf{X}\right) \approx \sum_{Z} \prod_{t=1}^{T} p\left ( z_{t}|z_{t-1}, W\right ) p\left (z_{t}|\mathbf{X}\right).
\label{eq:ctc_decode}
\end{equation}
We note that the tokens $W$ are embedded into $\mathbf{W}=(\mathbf{w}_{n} \in \mathbb{R}^{D} | n=1, \ldots, N)$ before being fed into the decoder module along with $\mathbf{H}$. The decoder then predicts the next token $w_{n}$ based on historical tokens $w_{1:n-1}$ and $\mathbf{H}$ via
\begin{equation}
\setlength{\abovedisplayskip}{4pt}
\setlength{\belowdisplayskip}{4pt}
  p\left(w_{n}|w_{1:n-1},\mathbf{X}\right)=\mathrm{Decoder}\left ( \mathbf{w}_{1:n-1}, \mathbf{H} \right ),
  \label{eq:asr_decoder}
\end{equation}
where $p(w_{n}|w_{1:n-1},\mathbf{X})$ is the posterior of $w_{n}$ given acoustic features and historical tokens, and $\mathrm{Decoder}(\cdot)$ denotes the Transformer decoder. The encoder-decoder module computes the token sequences by factorizing $p_{\mathrm{att}}\left(W|\mathbf{X}\right)$ as
\begin{equation}
  p_{\mathrm{att}}\left(W|\mathbf{X}\right) \approx \prod_{n=1}^{N} p\left (w_{n}|w_{1:n-1},\mathbf{X}\right).
\label{eq:att_decode}
\end{equation}

The model is optimized via a multi-task objective function
\begin{equation}
\setlength{\abovedisplayskip}{4pt}
\setlength{\belowdisplayskip}{4pt}
  \mathcal{L}_{\mathrm{asr}}=\alpha \mathcal{L}_{\mathrm{ctc}} + \left ( 1-\alpha  \right ) \mathcal{L}_{\mathrm{att}},
  \label{eq:loss_asr}
\end{equation}
where $\mathcal{L}_{\mathrm{ctc}}$ denotes the CTC loss, $\mathcal{L}_{\mathrm{att}}$ denotes the cross-entropy loss with label smoothing for the encoder-decoder branch~\cite{labelsmooth}, and $\alpha$ is a parameter associated with multi-task learning. The decoding process aims to maximize the linear combination of the logarithmic CTC and attention objectives such that the decoded token sequence is given by
\begin{equation}
\small
\setlength{\abovedisplayskip}{4pt}
\setlength{\belowdisplayskip}{4pt}
  \widehat{W} = \argmax_W \big \{ \alpha \mathrm{log} p_{\mathrm{ctc}}\left(W|\mathbf{X}\right) + \left ( 1-\alpha  \right ) \mathrm{log} p_{\mathrm{att}}\left(W|\mathbf{X}\right)\big\}.
  \label{eq:decoding}
\end{equation}
Here, $\widehat{W}$ is also referred to as a single hypothesis, and the final hypothesis of the given speech signal is chosen as the one with the highest likelihood among multiple hypotheses generated during the beam search. While the hybrid CTC/attention model has proven to be effective for the CS-ASR task, it performs CS-ASR similarly to monolingual ASR without exploiting any code-switching information, which consequently limits the CS-ASR performance.
\begin{figure}[t]
\setlength{\belowcaptionskip}{-1cm}
  \centering
  \includegraphics[width=0.8\linewidth]{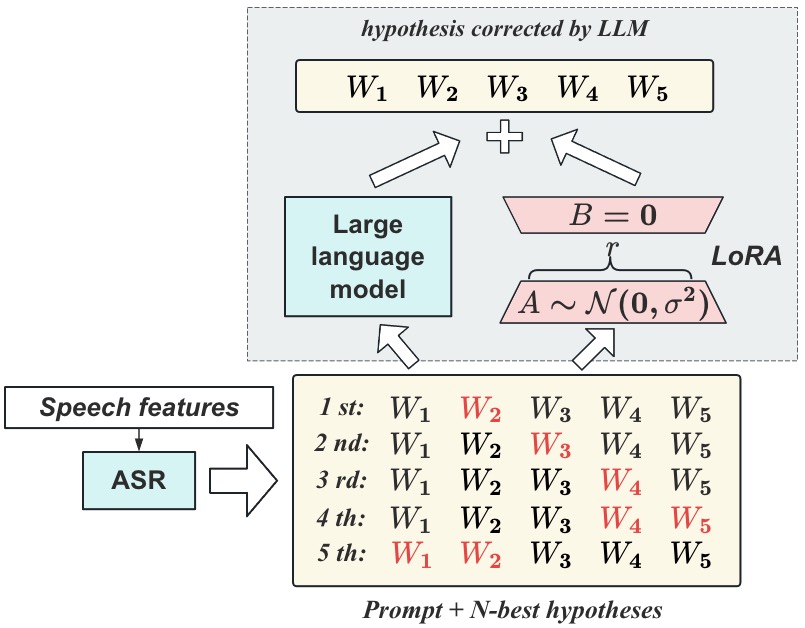}
  \caption{Generative error correction method for improving ASR with an external LLM proposed in~\cite{chen2023hyporadise}. Variable $r$ is the low intrinsic rank, and matrices $\mathbf{A}$ and $\mathbf{B}$ are initialized from a zero-mean Gaussian distribution and as zeros, respectively. The red and black tokens within the N-best hypotheses denote the wrongly and correctly predicted tokens, respectively.}
  \label{fig:lora}
\end{figure}

\subsection{Improving ASR via LLM and efficient fine-tuning}
\label{sec:pre_llm_lora}
Due to the scarcity of code-switching data, general code-switching language models moderately improve the performance of a CS-ASR system when incorporated through shallow fusion~\cite{liu23_icassp,liu2023enhancing}. Although developing an external language model on monolingual data or synthesized code-switching data is effective~\cite {peng2022internal,10023313, haibin2023}, the performance is limited by domain mismatch. Since large language models have achieved success in natural language processing and have been extended to computer vision and speech signal processing applications~\cite{gpt3, touvron2023llama, whisper}, we propose to adopt open-source LLMs, which are robust against diverse domains due to the large-scale training data, to improve CS-ASR by addressing complex token alternatives in bilingual scenarios. 

Recent works have also attempted to improve speech recognition through the use of LLMs. A direct method involves prompting an LLM using paired discrete speech and text embeddings~\cite{fathullah2023prompting}. A generative error correction method has also been applied to LLMs~\cite{chen2023hyporadise} as illustrated in Fig.~\ref{fig:lora}, where the final prediction is generated by summarizing and correcting the N-best ASR hypotheses~\cite{wang2021comprehensive}. This approach has shown effectiveness in monolingual ASR. However, code-switching leads to more token alternatives that have similar auditory or syntactic characteristics compared to a monolingual application\textemdash direct transference of the generative error correction to CS-ASR may not be desirable.

In addition, to fine-tune an LLM efficiently, low-rank adaptation~(LoRA)~\cite{hu2022lora} has been proposed. As shown in Fig.~\ref{fig:lora}, computational complexity is reduced by freezing the pre-trained LLM and injecting trainable rank decomposition matrices $\mathbf{A}$ and $\mathbf{B}$ into its Transformer-based layers. The forward pass is then defined as the linear combination of the pre-trained model $\mathbf{M}_0$ and the trained decomposed matrices $\mathbf{A}$ and $\mathbf{B}$ such that
\begin{equation}
\setlength{\abovedisplayskip}{4pt}
\setlength{\belowdisplayskip}{4pt}
  \left ( \mathbf{M}_0+\mathbf{\Delta M} \right ) \mathbf{X} = \left (\mathbf{M}_{0}+\mathbf{AB}  \right )\mathbf{X},
  \label{eq:lora}
\end{equation}
where $\mathbf{\Delta M}$ is the model parameters of the model update. Matrices $\mathbf{A}$ and $\mathbf{B}$ are initialized from a zero-mean Gaussian distribution and as zeros, respectively, such that $\mathbf{\Delta M}=\mathbf{AB}=0$ before training.

\section{Language Alignment Loss}
\label{sec:lal}
\begin{figure}[t]
  \centering
  \includegraphics[width=\linewidth]{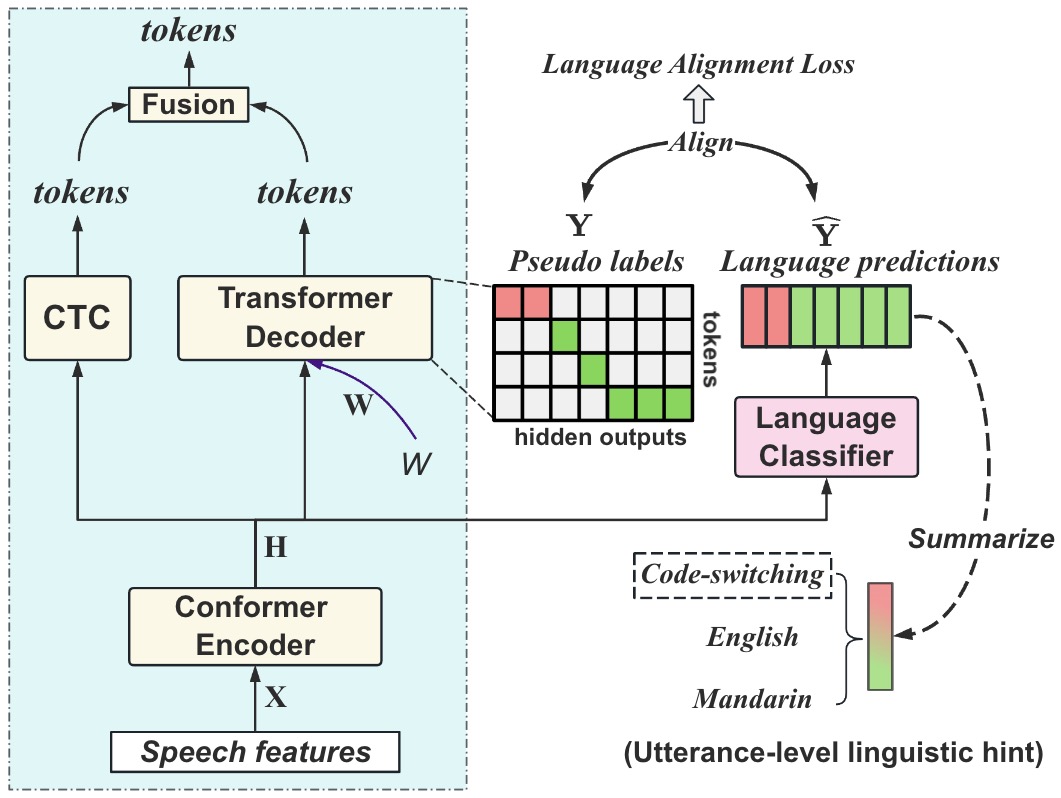}
  \caption{The hybrid CTC/attention model (in blue) with language alignment process.}
  \label{fig:asr_lal}
\end{figure}

Since multilingual ASR benefits from the supplementary language information offered by utterance-level one-hot language vectors, we propose to enhance CS-ASR performance by incorporating frame-level language information, enabling the detection of code-switching at high granularity. To this end, and as shown in Fig.~\ref{fig:asr_lal}, we introduce the LAL that is incorporated into the encoder-decoder framework for capturing language information.

\subsection{Frame-level language identification}
To capture frame-level language information, we employ a linear layer as a built-in language classifier. This layer takes the hidden output units of the encoder module as its input and aims to generate a language decision for each hidden output unit. 

Due to the lack of frame-level ground-truth or gold-standard language labels, existing works usually perform language identification in an unsupervised manner~\cite{bi_encoder,liu2023enhancing}. Nevertheless, this unsupervised frame-level classification extends beyond language identities and includes elements such as phonemic or domain information. Therefore, incorporating language information becomes important in guiding the unsupervised language identification process.

Although frame-level ground-truth language timestamps are unavailable, token-level language information can readily be inferred from text, particularly in cases where there is a notable contrast in character structure or morphology between the two languages. Past research has investigated the conversion of byte-pair encoding~(BPE) tokens into their respective language labels to facilitate language identification or diarization~\cite{jointrnnt, liu23_icassp, wang2023textderived}. Here, we employ a similar conversion strategy, where sub-tokens such as the BPE tokens are first transformed into token-level labels corresponding to their respective languages. The pseudo-frame-level language annotations are subsequently extracted by aligning frames to these token-level language labels.

\subsection{Aligning frames with token-level language labels}
\label{sec:align_f_to_t}
\begin{figure}[t]
  \centering
  \includegraphics[width=\linewidth]{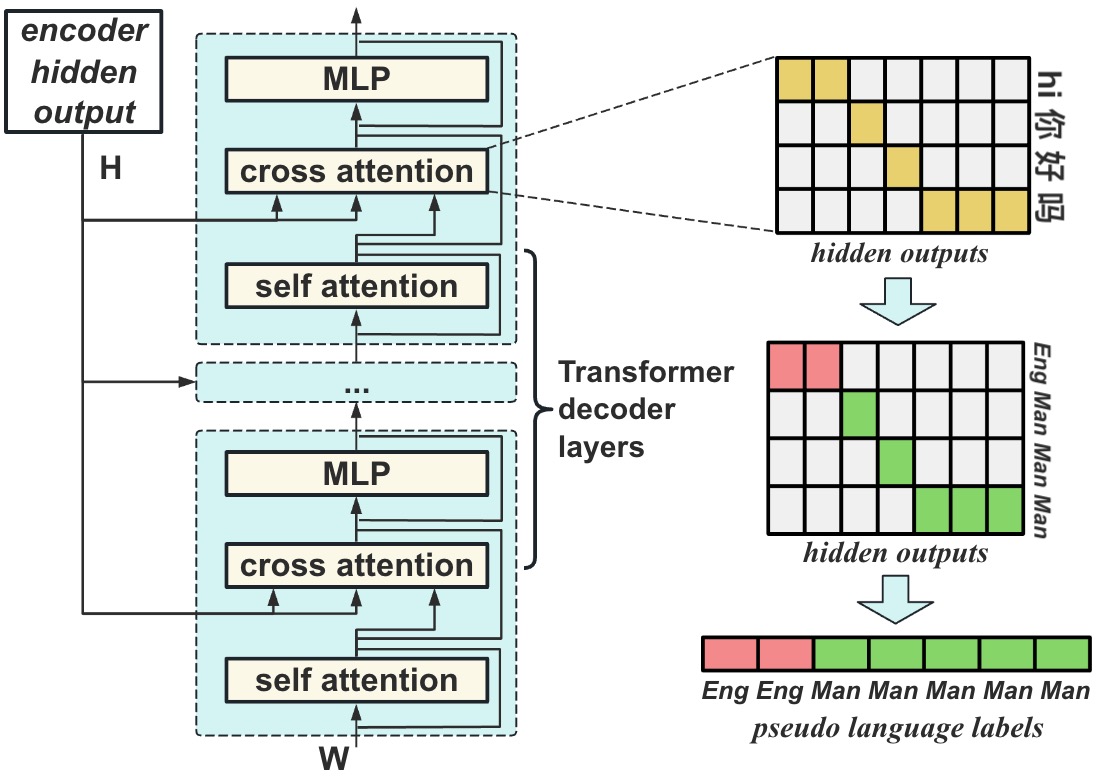}
  \caption{The pseudo frame-to-language label conversion process. On the right side, the upper matrix represents the cross-attention weight matrix averaged over all attention heads within the last decoder layer. The middle matrix represents the converted frame-to-language weight matrix and the pseudo-language labels are illustrated at the bottom. In this illustrative example, red and green cells denote the English and Mandarin tokens, respectively.}
  \label{fig:pseudo_lang}
\end{figure}

The encoder-decoder model achieves ASR by mapping speech features to tokens. The alignment between speech frames (i.e., encoder hidden outputs) and text (i.e., tokens) is inherently learned through the cross-attention process illustrated on the left of Fig.~\ref{fig:pseudo_lang}. Given multiple attention heads within the last Transformer decoder layers, their cross-attention matrices are averaged to generate the weight matrix $\mathbf{Atten} \in \mathbb{R}^{T^{\prime} \times N}$ of the speech-to-text alignment shown in the top matrix of Fig.~\ref{fig:pseudo_lang}. The alignment between hidden outputs and languages can then be derived from the frame-to-token weight matrix as illustrated from the top to bottom matrices in Fig~\ref{fig:pseudo_lang}. Specifically, the averaged cross-attention weight matrix can be decomposed into vectors along $T^{\prime}$, being $\mathbf{Atten} = (\mathbf{atten}_{t} \in \mathbb{R}^{N}| t=1, \ldots, T^{\prime})$. Each element in $\mathbf{atten}_{t}$ denotes the attention weight corresponding to a BPE token within the input token sequence. Each frame can then be assigned a pseudo-language label, which is the language of the BPE token corresponding to the highest weight. This is achieved via
\begin{equation}
    \mathbf{y}_{t} = \mathrm{T2L}\big ( \argmax_{n} \left(\mathbf{atten}_{t}\right) \big),
  \label{eq:token_to_lang}
\end{equation}
where $\mathrm{T2L}(\cdot)$ represents the conversion from the $n$-th BPE token with the highest attention weight in $\mathbf{atten}_{t}$ to its language label.

The pseudo-language labels are concatenated to form a sequence $\mathbf{Y}=(\mathbf{y}_{t} \in \mathbb{R}^{C}| t=1, \ldots, T^{\prime})$. Here, $T^{\prime}$ is the number of hidden output units and $C$ is the number of languages, where all special tokens are treated as a third language in addition to the two target languages. Each hidden output unit is subsequently projected to a language decision $\mathbf{y}_{t}$ by a language classifier comprising one linear layer
\begin{equation}
    \mathbf{\widehat{y}}_{t} = \mathrm{Linear}\left ( \mathbf{h}_{t} \right ),
  \label{eq:lang_layer}
\end{equation}
where $\mathrm{Linear}\left ( \cdot \right )$ denotes computations within a linear layer. Defining $\mathrm{exp}(\cdot)$ as the exponential operation, the language alignment loss for each speech sample is computed via a cross-entropy function
\begin{equation}
    \mathcal{L}_{\mathrm{lid}} = \frac{-1}{T^{\prime}}\sum_{t=1}^{T^{\prime}}\sum_{c=1}^{C}\mathrm{log}\frac{\mathrm{exp}\left ( \mathbf{\widehat{y}}_{t,c} \right )}{\sum_{i=1}^{C}\mathrm{exp}\left (\mathbf{\widehat{y}}_{t,i} \right )}\mathbf{y}_{t,c},
  \label{eq:lang_loss_1}
\end{equation}
where $\mathbf{\widehat{y}}_{t,i}$ denotes the value at the $i$th dimension of the language prediction vector for the $t$th hidden output vector. Similarly, $\mathbf{y}_{t,c}$ denotes the value at the $c$th dimension of the one-hot pseudo-language label vector for the $t$th hidden output vector such that $\mathbf{y}_{t,c}=1$ for the target language with the remaining elements being zero. The proposed LAL is employed on the encoder representations, enriching the ASR encoder (acoustic end) with frame-level language information. The ASR decoder is not explicitly supervised by this objective function.

The key distinction between the proposed LAL and general LID lies in the ability of the former to refine pseudo-labels iteratively during training, enabling a mutual optimization process between ASR and LID tasks. The performance of the CS-ASR model improves during model training as the language information is incorporated, reaching its peak performance upon being optimized completely. The improved ASR performance contributes to increasingly accurate language labels. This improvement, in turn, leads to higher language identification performance. Therefore, extracting pseudo-language labels and language identification has become integral to the iterative optimization process during training.

\subsection{Balancing training via token-level language weights}
The grammatical structure of code-switching data defines the matrix language as the main language and the embedded language as the secondary language~\cite{myers1997duelling}. While code-switching corpora such as SEAME do not exhibit a dominant language~\cite{seame, syntactic_LM}, the matrix language may prevail in some code-switching corpora, resulting in an imbalanced distribution of tokens and speech frames between the two languages~\cite{shi2020asru}. Therefore, a CS-ASR model trained on such imbalanced data may overfit the matrix language and underfit the embedded language.

While computing the number of speech frames for each language is challenging, the token distribution within a code-switching corpus can be assessed before training. Therefore, we propose to address the imbalance issue by incorporating language weights in (\ref{eq:lang_loss_1}), resulting in a weighted cross-entropy function
\begin{equation}
    \mathcal{L}_{\mathrm{lal}} = \frac{-1}{T^{\prime}}\sum_{t=1}^{T^{\prime}}\sum_{c=1}^{C}w^{\mathrm{lang}}_{c} \mathrm{log}\frac{\mathrm{exp}\left ( \mathbf{\widehat{y}}_{t,c} \right )}{\sum_{i=1}^{C}\mathrm{exp}\left (\mathbf{\widehat{y}}_{t,i} \right )}\mathbf{y}_{t,c},
  \label{eq:lang_loss_2}
\end{equation}
where $w^{\mathrm{lang}}_{c}$ denotes the normalized language weight of language $c$ and is inversely proportional to its token count within the training data. However, the token ratio may not align precisely with speech frames due to variations in speech rates across languages. We therefore propose to tune them initially based on the token ratio and accounting for speech rate variations when applying language weights to the LAL. The relationship between the initial language weights is given by
\begin{equation}
    \frac{w^{\mathrm{lang}}_{\mathrm{Eng}}}{w^{\mathrm{lang}}_{\mathrm{Man}}} \propto \frac{\mathrm{Count}(\mathrm{tokens}_{\mathrm{Man}})}{\mathrm{Count}(\mathrm{tokens}_{\mathrm{Eng}})},
  \label{eq:initial_weight}
\end{equation}
where $\mathrm{Count}(\cdot)$ denotes the number of BPE tokens belonging to the language. In particular, fewer tokens in the training data and a higher speech rate result in a higher language weight.

With the above, the CS-ASR model is optimized via an objective function being the sum of CTC loss, encoder-decoder loss, and the proposed LAL such that
\begin{equation}
  \mathcal{L}_{\mathrm{asr}}=\alpha \mathcal{L}_{\mathrm{ctc}} + \left ( 1-\alpha  \right ) \mathcal{L}_{\mathrm{att}} + \beta \mathcal{L}_{\mathrm{lal}},
  \label{eq:loss_all}
\end{equation}
where $\beta$ denotes the weight of the language alignment loss during training. The decoding process is similar to (\ref{eq:decoding}).

\subsection{Linguistic hint for prompting LLM}
\label{sec:hint}
\begin{figure}[t]
  \centering
  \includegraphics[width=\linewidth]{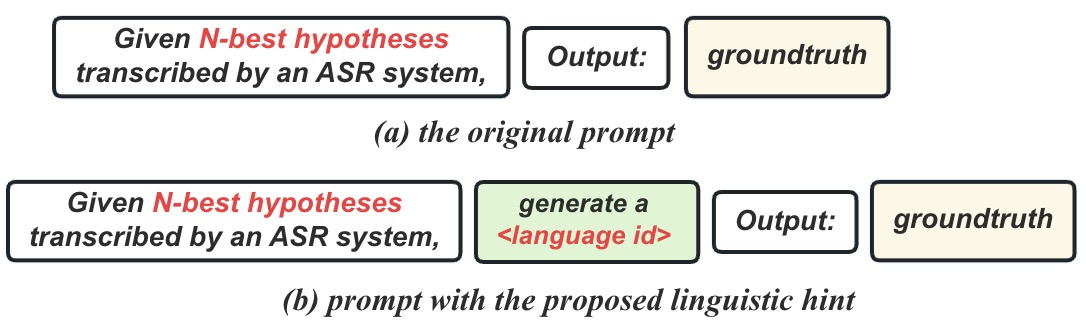}
  \caption{The original (a) and proposed (b) prompts during fine-tuning LLM and performing generative error correction, \textit{language id} is selected from two languages (``English" and ``Mandarin" in this work) and ``multilingual".}
  \label{fig:prompt}
\end{figure}

With reference to~Fig.~\ref{fig:lora}, we propose to adopt the LLM-based generative error correction method to improve CS-ASR~\cite{chen2023hyporadise}. To fine-tune the LLM efficiently, LoRA is employed while the LLM is kept frozen during training, as illustrated in Section~\ref{sec:pre_llm_lora}. The prompt originally designed in~\cite{chen2023hyporadise} is used as shown in Fig.~\ref{fig:prompt}(a), where the N-best hypotheses are extracted from the ASR output before being inserted in the prompt. The LLM is then optimized to deduce the correct transcript, which is the ground-truth transcription shown in Fig.~\ref{fig:prompt} during training, by leveraging the information provided in these hypotheses. However, code-switching gives rise to more intricate token alternatives with similar auditory or syntactic characteristics compared to a monolingual application. This complexity persists when performing generative error correction for ASR hypotheses. Inspired by the use of chain-of-thought as additional supervision when fine-tuning LLM for downstream tasks~\cite{wei2022chain, hsieh-etal-2023-distilling}, we propose to employ an additional linguistic hint during prompting to address the aforementioned challenge in CS-ASR.

To this end, various methods for linguistic hint extraction can be used. An acoustic-biased linguistic hint can be derived from the by-product of the proposed LAL. As shown in Fig.~\ref{fig:asr_lal}, frame-level language predictions are first normalized using the softmax function before generating an utterance-level language decision. In the context of this work, this decision can either be monolingual or multilingual, with the former providing a single language code. In addition, the linguistic hint can be obtained from the decoded hypotheses (i.e., text information), allowing it to be employed in conjunction with the acoustic-biased hint through the weighted voting mechanism.

The linguistic hint is then inserted into the used prompt during fine-tuning as shown in Fig.~\ref{fig:prompt}(b). We propose two types of hints\textemdash the monolingual hint, where only $<$\textit{language~id}$>$ words are included in the transcription, and the code-switching hint, where speech is multilingual and that $<$\textit{both language ids}$>$ words are included in the transcription. 

\section{Dataset, experiment, and model configuration}
\label{sec:data_exp_model}

\subsection{Datasets}
\label{sec:datasets}
We conducted the experiments on data extracted from the ASRU 2019 Mandarin-English code-switching speech recognition challenge and the SEAME dataset~\cite{seame, shi2020asru}. The ASRU 2019 Mandarin-English code-switching speech recognition challenge consists of four datasets, including a 500-hour Mandarin-only training set, a 200-hour intra-sentence English-Mandarin code-switching training set, a 40-hour intra-sentence English-Mandarin code-switching development set, and a 20-hour intra-sentence English-Mandarin code-switching test set. In the experiments, the models were trained on the 200-hour CS training set, validated on the development set, and evaluated on the test set. The SEAME dataset, on the other hand, is a Mandarin-English code-switching corpus containing spontaneous conversational speech~\cite{seame}. This dataset encompasses both intra- and inter-sentence code-switching speech. We divided the SEAME dataset into a 96.6-hour training set, a 4.9-hour validation set, and two test sets denoted by $\text{test}_{\texttt{man}}$ and $\text{test}_{\texttt{sge}}$ following the same partitioning method described in \cite{zeng19_interspeech}. Detailed information regarding the test sets is provided in Table~\ref{tab:data}. Table~\ref{tab:seame} highlights the duration ratio of each language, where the language labels are annotated at the utterance level. Example text sentences from the two datasets are shown in Table~\ref{tab:where_error_happen}.

While both datasets involve English-Mandarin code-switching, the primary distinction between them lies in the accent. The ASRU dataset, recorded in mainland China, is characterized by a dominant Chinese accent and text. In the training and development sets, each sentence, on average, consists of 8.6 Chinese characters and 1.6 English words. In contrast, the SEAME dataset comprises audio recordings from Singapore and Malaysia, featuring South-East Asian accents. On average, sentences in the SEAME training and development sets contain 9.5 Chinese characters and 4.4 English words. Additionally, code-switching occurs more frequently within the SEAME dataset compared to the ASRU data due to the bilingual education and language policies in Singapore and Malaysia~\cite{dixon2005bilingual}. This suggests that the SEAME data might pose greater challenges for CS-ASR compared to the ASRU data.

\begin{table}[t]
\centering
\caption{Details of two datasets in terms of division and durations}
\label{tab:data}
\renewcommand{\arraystretch}{1}
\setlength{\tabcolsep}{7mm}{
\begin{tabular}{c|c|c}
\toprule
\textbf{Corpus}                 & \textbf{Subset}  & \textbf{Duration (hours)} \\ \midrule
\multirow{3}{*}{ASRU}  & train  & 193.0            \\ 
                       & dev    & 21.3             \\
                       & test   & 20.4             \\ \midrule
\multirow{4}{*}{SEAME} & train  & 96.6             \\ 
                       & dev    & 4.9              \\ 
                       & $\text{test}_{\texttt{man}}$  & 7.5 \\ 
                       & $\text{test}_{\texttt{sge}}$  & 3.9 \\ \bottomrule
\end{tabular}}
\end{table}

\begin{table}[t]
\centering
\caption{Utterance-level duration ratios and dataset-level token distribution (English words and Mandarin characters) ratios in the ASRU and SEAME test sets}
\label{tab:seame}
\renewcommand{\arraystretch}{1}
\setlength{\tabcolsep}{3.5mm}{
\begin{tabular}{c|ccc|cc}
\toprule
\multirow{2}{*}{\textbf{Subset}}  & \multicolumn{3}{c|}{\textbf{Duration ratio (\%)}} & \multicolumn{2}{c}{\textbf{Token ratio (\%)}}  \\ 
          & Man & Eng & CS & Man & Eng \\ \midrule
ASRU test                               & 0   & 0  & 100 & 89 & 11\\ \midrule
$\text{SEAME test}_{\texttt{man}}$       & 14  & 7  & 79  & 74 & 26\\ 
$\text{SEAME test}_{\texttt{sge}}$       & 6   & 41 & 53  & 37 & 63\\ \bottomrule
\end{tabular}}
\end{table}

\subsection{Data preprocessing}
Since the SEAME dataset contains a small training set of approximately 98~hrs, we augmented the training data using speed perturbation and SpecAugment~\cite{pertub, specaug} for training models from scratch. Two training strategies were adopted, wherein one develops the model on data without speed perturbation while the other trains the model on the entire augmented data. The speech perturbation was applied with factors 0.9, 1.0, and 1.1. With respect to the ASRU dataset, only SpecAugment was applied for data augmentation. SpecAugment adopted the default setup in ESPnet for two datasets~\cite{espnet}. The time-warp mask size was set to five, and two time and frequency masks were applied, with their lengths uniformly selected from the range of 0 to 40 for time masks and 0 to 30 for frequency masks. Speech samples within both corpora were segmented into durations ranging from 0.1 to 20~s. We extracted $F=80$ dimensional log-Mel-Fbank features for each speech segment before applying the cepstral mean and variance normalization.

For the train-from-scratch models, we employed BPE to tokenize the English words in the two English-Mandarin code-switching corpora and split all Mandarin words into individual characters. For the SEAME dataset, this resulted in a total of $V=5,628$ tokens, including 3,000 English BPE tokens, 2,624 Mandarin characters, and four special tokens (\textit{$<$unk$>$}, \textit{$<$noise$>$}, \textit{$<$blank$>$}, and \textit{$<$sos/eos$>$}). For the ASRU data, the same tokenization process yielded a total of $V=6,923$ tokens comprising 3,000 English BPE tokens, 3,920 Mandarin characters, and three special tokens (\textit{$<$unk$>$}, \textit{$<$blank$>$}, and \textit{$<$sos/eos$>$}). The variable $C=3$ denotes three language classes converted from special, English, and Mandarin tokens.

We fine-tuned the Chinese LLaMA-2 on a subset of the SEAME training set comprising approximately 60,000 speech segments and the development set of the ASRU data comprising approximately 20,000 speech segments. N-best lists of the above speech segments were subsequently extracted and incorporated into the prompts. In addition, we removed the \textit{$<$noise$>$} and \textit{$<$unk$>$} labels from the N-best list since the LLM can hardly address these special tokens.

For experiments involving Whisper models, we adopted the data preprocessing and tokenization methods outlined in \cite{whisper}. Forced alignment is utilized to generate frame-level language annotations and is performed using the fine-tuned Whisper-small (FA-LB) and Whisper-large (FA-UB) models.

\begin{figure}[t]
  \centering
  \includegraphics[width=\linewidth]{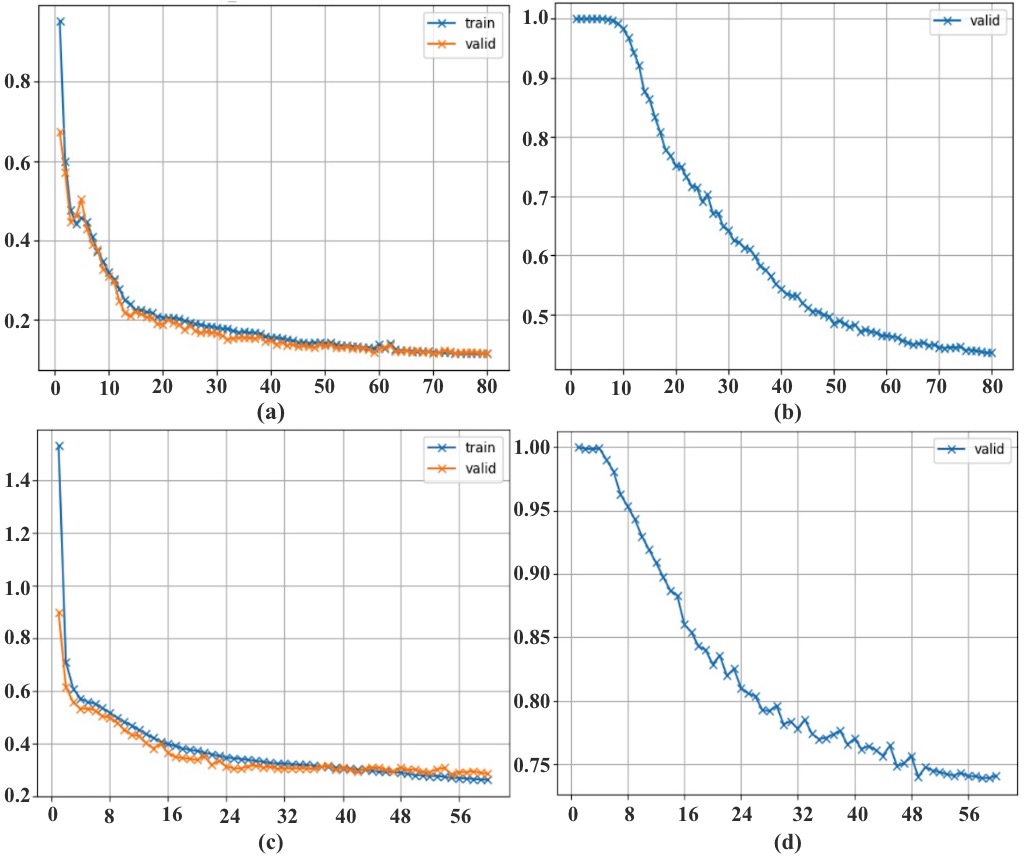}
  \caption{The convergence of the proposed language alignment loss (y-axis for the left figures) during training and validation on the (a) ASRU data and (c) SEAME data, and the corresponding MER (y-axis for the right figures) on the validation set of (b) the ASRU data and (d) SEAME data, against training epochs (x-axis).}
  \label{fig:loss_wer}
\end{figure}

\subsection{Model configuration and experimental setup}
The baseline model is a Conformer-based hybrid CTC/Attention ASR model comprising twelve Conformer encoder layers, six Transformer decoder layers, and a CTC module. The Conformer employs the macaron structure~\cite{conformer}, where its convolutional neural network~(CNN) module has a kernel size of 15. The swish activation function is applied~\cite{swish}. A CNN layer first subsamples the input features and projects them into $D=256$ dimensions before feeding into the macaron modules. All attention layers within the encoder and decoder modules have four attention heads with input and output dimensions being $D=256$, and the inner layer of the position-wise feed-forward network is of dimension 2048.

The bi-encoder and language posterior bias methods have also been implemented as baselines~\cite{bi_encoder, liu23_icassp}. In the bi-encoder CS-ASR system, we replicated the encoder module of the baseline model described in the previous paragraph to form the bi-encoder module. The outputs of these two encoders are fed into a linear layer, which serves as a built-in LID module with an output dimension of two (corresponding to two languages), before calculating the weighted encoder outputs. It is useful to note that the encoder modules of the bi-encoder method are pre-trained on English and Mandarin corpora separately. Nevertheless, this pre-training was deliberately omitted to ensure a fair comparison. For the language posterior bias-based~(LPB) CS-ASR model, the model configuration recommended in~\cite{liu23_icassp} was used. The LPB system closely resembles the baseline hybrid CTC/attention model, with an additional language diarization decoder. This language diarization decoder mirrors the structure of the ASR decoder.

We trained the baseline model on the ASRU and SEAME data for seventy and fifty epochs, respectively. The proposed method and reproduced works were trained for an additional ten epochs due to their more complex objective functions or parameters. All models were optimized by an Adam optimizer on two RTX 3090 GPUs. The learning rate increased from 0 to 0.001 over 25,000 update steps, followed by a cosine annealing decay. Parameter $\alpha=0.3$ and a label smoothing factor of 0.1 were used in~(\ref{eq:loss_asr}) and~(\ref{eq:loss_all}). The ten best models during validation were averaged for inference. We adopted the ten-best beam search method with $\alpha=0.4$ in~(\ref{eq:decoding}). 

Experiments associated with the proposed linguistic hint utilized a Chinese LLaMA-2-7B model\footnote{https://huggingface.co/hfl/chinese-llama-2-7b}. The 5-best hypotheses were first extracted from the ASR output before being used in the prompt to fine-tune the LLM. Fine-tuning was performed via LoRA with rank $r=4$ and the AdamW optimizer for ten epochs with a batch size of 128 on the SEAME data, where LoRA is applied to the query and value modules within the self-attention modules. The learning rate was increased from 0 to 0.0002 over 100 update steps, followed by a linear decay. During inference, a temperature of 0.7 was applied to allow for the creativity of the LLM, while other hyper-parameters were fixed at their default settings.

Experiments relevant to Whisper models utilized the Whisper-small model~\cite{whisper}. Fine-tuning was conducted using an AdamW optimizer over five epochs, with a learning rate increased from 0 to $1 \times 10^{-5}$ over 10,000 update steps, followed by a linear decay. The training employed a batch size of eight samples, with gradients accumulated every two updates. Inference was carried out using the best-performing model on the development sets, with a beam size of one. Language prompts \textit{$<$zh$>$} and \textit{$<$en$>$} were used for the ASRU and SEAME data, respectively. Since Whisper models were pre-trained on large-scale data, $\beta$ was set to 0.05 and 0.01 for ASRU and SEAME data, respectively, to prevent the LAL or frame-level LID from exerting excessive influence during fine-tuning. The frame-to-token alignment for LAL is computed via the cross-attention matrices within the last decoder layer. LoRA was applied to the query, key, value, and linear modules with rank $r=8$ during parameter-efficient fine-tuning.

Evaluation of CS-ASR systems was quantified via the mix error rate~(MER) comprising word error rate~(WER) for English and character error rate~(CER) for Mandarin. LID accuracy was calculated at the utterance level.
\section{Results and analysis}
\label{sec:results}

\subsection{Impact of $\beta$ values}
To assess the impact of $\beta$ on the CS-ASR performance for the proposed LAL, we first compare the performance of the hybrid CTC/attention model optimized with LAL via~(\ref{eq:loss_all}) for various $\beta$ values on the ASRU data. Results summarized in Table~\ref{tab:asru_results} highlight that with the incorporation of LAL during training, the hybrid CTC/attention model consistently outperforms the vanilla model. Notably, the best MER of 11.9\% is observed when $\beta=1.5$, achieving a relative improvement of 7.03\% compared to the vanilla model. In addition, this performance improvement remains consistent across various LAL parameters, particularly within the range of $1.0\leq\beta\leq 3.0$.

\begin{table}[t]
    \centering
\caption{Performance evaluation of the proposed method on the ASRU data with different $\beta$ values in terms of substitutions, deletions, insertions, and the total MER (\%)}
\label{tab:asru_results}
\renewcommand{\arraystretch}{1.1}
\setlength{\tabcolsep}{1.9mm}{
\begin{tabular}{c|c|ccc|c}
\toprule
\textbf{Method} &  \textbf{$\beta$ for LAL}& \textbf{Sub} $\downarrow$ & \textbf{Del} $\downarrow$&  \textbf{Ins} $\downarrow$ & \textbf{MER} $\downarrow$\\ \midrule
Hybrid CTC/atten&  0&  11.5&   0.6& 0.7& 12.8\\ \midrule
        & 0.5& 11.2& 0.6& 0.7& 12.4\\ 
        & 1.0& 10.9& 0.6& 0.7& 12.1\\ 
        & 1.5& 10.7& 0.5& 0.6&\textbf{11.9}\\ 
 + LAL  & 2.0& 10.8& 0.5& 0.6& 12.0\\ 
        & 2.5& 10.9& 0.5& 0.6& 12.0\\ 
        & 3.0& 10.9& 0.5& 0.6& 12.0\\ 
        & 4.0& 11.4& 0.6& 0.6& 12.6\\ 
        & 5.0& 11.0& 0.6& 0.6& 12.2\\ \bottomrule
\end{tabular}}
\end{table}

We next conduct experiments on the SEAME dataset without using speed perturbation during training and present the results in Table~\ref{tab:seame_results}. Similar to the ASRU dataset, the hybrid CTC/attention model that incorporates the proposed LAL outperforms the vanilla configuration. In addition, the hybrid CTC/attention model with LAL exhibits consistently high performance for $1.0\leq\beta\leq 3.0$. This robustness on various datasets and $\beta$ values indicates the effectiveness of the proposed LAL. As opposed to results achieved on the ASRU dataset, the highest overall performance on the SEAME data is achieved when $\beta=0.1, 0.5,$ and 3.0. Due to the different duration ratios of languages of the two test sets, the above implies that a lower $\beta$ value leads to a higher performance on the dataset containing predominantly monolingual data. 

The impact of $\beta$ values can also be discussed from the perspective of accent. Compared to the ASRU data, which is characterized by a dominant Chinese accent, the SEAME data is characterized by Southeast Asian accents. Due to the impact of Southeast Asian accents on the two languages, the English and Mandarin speech within the SEAME dataset can exhibit auditory similarity, making LID for the SEAME data more challenging than that for the ASRU data. Therefore, a larger $\beta$ value is required to achieve high performance when training the CS-ASR model on the SEAME data.
\begin{table}[t]
\centering
\caption{Performance evaluation of the proposed method trained without speed perturbation on the SEAME dataset with different $\beta$ values in terms of substitutions, deletions, insertions, and the total MER (\%)}
\label{tab:seame_results}
\renewcommand{\arraystretch}{1.1}
\setlength{\tabcolsep}{1.0mm}{
\begin{tabular}{c|c|c|ccc|c}
\toprule
\textbf{Method} &  \textbf{$\beta$ for LAL} & \textbf{Subset} & \textbf{Sub} $\downarrow$ & \textbf{Del} $\downarrow$&  \textbf{Ins} $\downarrow$ & \textbf{MER} $\downarrow$\\ \midrule
\multirow{2}{*}{Hybrid CTC/atten} & \multirow{2}{*}{0}  & $\text{test}_{\texttt{man}}$  & 12.0 & 3.0 & 2.2 & 17.2 \\ 
                                  &                     & $\text{test}_{\texttt{sge}}$  & 17.3 & 4.1 & 3.1 & 24.5 \\ \midrule
\multirow{7}{*}{+ LAL}      & \multirow{2}{*}{0.1} & $\text{test}_{\texttt{man}}$ & 11.8 & 2.9 & 2.2 & 16.8 \\ 
                            &                      & $\text{test}_{\texttt{sge}}$ & 17.0 & 4.0 & 3.0 & \textbf{23.9} \\ \cmidrule{2-7}
                            & \multirow{2}{*}{0.5} & $\text{test}_{\texttt{man}}$ & 11.8 & 2.9 & 2.2 & 16.8 \\ 
                            &                      & $\text{test}_{\texttt{sge}}$ & 17.0 & 4.0 & 3.0 & \textbf{23.9} \\ \cmidrule{2-7}
                            & \multirow{2}{*}{1.0} & $\text{test}_{\texttt{man}}$ & 11.7 & 3.0 & 2.1 & 16.8 \\ 
                            &                      & $\text{test}_{\texttt{sge}}$ & 17.1 & 4.0 & 3.1 & 24.1 \\ \cmidrule{2-7}
                            & \multirow{2}{*}{2.0} & $\text{test}_{\texttt{man}}$ & 11.9 & 3.0 & 2.0 & 16.9 \\ 
                            &                      & $\text{test}_{\texttt{sge}}$ & 17.0 & 4.0 & 3.1 & 24.1 \\ \cmidrule{2-7}
                            & \multirow{2}{*}{3.0} & $\text{test}_{\texttt{man}}$ & 11.7 & 3.0 & 2.0 & \textbf{16.7} \\ 
                            &                      & $\text{test}_{\texttt{sge}}$ & 17.0 & 4.0 & 3.0 & 24.0 \\ \cmidrule{2-7}
                            & \multirow{2}{*}{4.0} & $\text{test}_{\texttt{man}}$ & 11.9 & 3.0 & 2.0 & 16.9 \\ 
                            &                      & $\text{test}_{\texttt{sge}}$ & 17.1 & 4.2 & 3.0 & 24.2 \\ \cmidrule{2-7}
                            & \multirow{2}{*}{5.0} & $\text{test}_{\texttt{man}}$ & 11.8 & 3.0 & 2.2 & 17.0 \\ 
                            &                      & $\text{test}_{\texttt{sge}}$ & 17.4 & 4.1 & 3.1 & 24.5 \\ \bottomrule
\end{tabular}}
\end{table}

\begin{table}[t]
    \centering
\caption{Performance of the proposed method with $\beta=1.5$ and different language weights to balance the Mandarin-dominant ASRU data during training in terms of English WER, Mandarin CER, and MER (\%). ``$ft$" denotes fine-tuning}
\label{tab:asru_balancing}
\renewcommand{\arraystretch}{1.1}
\setlength{\tabcolsep}{3mm}{
\begin{tabular}{ccc|ccc}
\toprule
\multicolumn{3}{c|}{\textbf{Language weights ($\beta=1.5$})} & \multicolumn{3}{c}{\textbf{ASRU test}}     \\ \midrule
\textbf{Other} & \textbf{Eng} & \textbf{Man} & \textbf{Eng} $\downarrow$ & \textbf{Man} $\downarrow$ & \textbf{MER} $\downarrow$ \\ \midrule
\multicolumn{3}{c|}{vanilla (all 1)}                       & 35.4         & 9.3          & 11.9         \\ \midrule
1              & 10           & 1            & 35.2         & 9.3          & 11.8         \\ 
1              & 50           & 1            & 35.3         & 9.3          & 11.8         \\ 
1              & 100          & 1            & \textbf{35.1}& \textbf{9.2} & \textbf{11.7} \\ 
1              & 100          & 100          & 35.8         & 9.5          & 12.1         \\ 
1              & 1000         & 1            & 35.4         & 9.3          & 11.8         \\ \bottomrule
\end{tabular}}
\end{table}

\subsection{Balancing the ASRU dataset during training}
\label{sec:balance}

The effect of language weights employed in (\ref{eq:lang_loss_2}) for balancing the Mandarin-dominant ASRU data is shown in Table~\ref{tab:asru_balancing}. We note that, similar to the number of English frames, the number of frames for the class ``other" is also significantly lower than that for the Mandarin frames. Since the class ``other'' does not contribute to the language identities in the CS-ASR task, this class is not balanced during training and the weight is always set to $w_{other}^{\mathrm{lang}}=1$ except for the learnable language weight. In this work, $w_c^{\mathrm{lang}}=1$ is the default setup for all classes.

Results presented show that the CS-ASR model gains moderate performance improvement on both English and Mandarin data from high English weights, where the weights of ``other" and ``Mandarin" were set to 1. This is consistent with our assumption that a high English weight can achieve balance for the secondary language during training, which consequently improves the model performance. Moreover, the highest performance is achieved when the weight of English is set to 100. It is also worth noting that the ratio between English to Mandarin weights is higher than the token ratio due to the difference between their speech rates. 

\begin{table}[t]
\centering
\caption{Results of the proposed method on SEAME dataset after employing speed perturbation with factor 0.9, 1.0, and 1.1 during training in terms of substitutions, deletions, insertions, and the total MER (\%)}
\label{tab:seame_results_perturb}
\renewcommand{\arraystretch}{1.1}
\setlength{\tabcolsep}{1.0mm}{
\begin{tabular}{c|c|c|ccc|c}
\toprule
\textbf{Method} &  \textbf{$\beta$ for LAL} & \textbf{Subset} & \textbf{Sub} $\downarrow$ & \textbf{Del} $\downarrow$&  \textbf{Ins} $\downarrow$ & \textbf{MER} $\downarrow$\\ \midrule
\multirow{2}{*}{Hybrid CTC/atten} & \multirow{2}{*}{0}  & $\text{test}_{\texttt{man}}$  & 11.5 & 3.0 & 2.0 & 16.6 \\ 
                                  &                     & $\text{test}_{\texttt{sge}}$  & 16.4 & 3.9 & 3.0 & 23.3 \\ \midrule
\multirow{6}{*}{+ LAL}      & \multirow{2}{*}{0.5} & $\text{test}_{\texttt{man}}$ & 11.6 & 3.0 & 2.1 & 16.7 \\
                            &                      & $\text{test}_{\texttt{sge}}$ & 16.4 & 4.1 & 3.0 & 23.5 \\ \cmidrule{2-7}
                            & \multirow{2}{*}{1.0} & $\text{test}_{\texttt{man}}$ & 11.6 & 2.9 & 2.2 & 16.7 \\
                            &                      & $\text{test}_{\texttt{sge}}$ & 16.4 & 4.0 & 3.1 & 23.6 \\ \cmidrule{2-7}
                            & \multirow{2}{*}{2.0} & $\text{test}_{\texttt{man}}$ & 11.5 & 3.0 & 2.0 & 16.5 \\
                            &                      & $\text{test}_{\texttt{sge}}$ & 16.5 & 4.1 & 3.0 & 23.5 \\ \cmidrule{2-7}
                            & \multirow{2}{*}{3.0} & $\text{test}_{\texttt{man}}$ & 11.3 & 3.0 & 2.1 & \textbf{16.4} \\ 
                            &                      & $\text{test}_{\texttt{sge}}$ & 16.2 & 4.0 & 2.9 & \textbf{23.3}\\ \cmidrule{2-7}
                            & \multirow{2}{*}{4.0} & $\text{test}_{\texttt{man}}$ & 11.7 & 3.1 & 2.1 & 16.8 \\
                            &                      & $\text{test}_{\texttt{sge}}$ & 16.6 & 3.9 & 3.0 & 23.6 \\ \cmidrule{2-7}
                            & \multirow{2}{*}{5.0} & $\text{test}_{\texttt{man}}$ & 11.6 & 3.0 & 2.1 & 16.7 \\
                            &                      & $\text{test}_{\texttt{sge}}$ & 16.6 & 4.0 & 3.1 &  23.6\\ \bottomrule
\end{tabular}}
\end{table}

\subsection{Impact of speed perturbation}
As discussed in Section~\ref{sec:balance}, the frame-level language identification can be affected by speech rate. We, therefore, investigate the impact of the speed perturbation on the CS-ASR performance when LAL is incorporated. The experiments were conducted on the SEAME dataset and in contrast to the experiments shown in Table~\ref{tab:seame_results}, we utilize speed perturbation to augment the training data with results summarized in Table~\ref{tab:seame_results_perturb}.

These results show that the hybrid CTC/attention model achieves higher performance after employing speed perturbation as data augmentation. The proposed method achieves the highest performance on the ASRU dataset when $\beta=1.5$. Since a higher $\beta$ value is required for the SEAME data to achieve the highest performance, this also underpins that the SEAME dataset presents greater challenges in language discrimination than the ASRU dataset. While speed perturbation is effective for data augmentation and overall ASR performance improvement, it introduces inconsistencies in the alignment between text tokens and acoustic frames, where a single token may correspond to different numbers of frames under varying speech rates. These inconsistencies can reduce the reliability of frame-level supervision and the effectiveness of the proposed LAL. Therefore, further incorporation of the proposed LAL may lead to lower performance. Notwithstanding the above, LAL can still yield moderate performance improvement under speed perturbation when appropriately tuned, where the hybrid CTC/attention model with LAL being incorporated achieves the highest performance when $\beta=3$. This is consistent with the performance of its counterpart on the SEAME dataset without speed perturbation. Therefore, although the overall improvement may be less significant compared to training without perturbation, LAL remains beneficial and applicable in real-world CS-ASR scenarios, particularly when speed perturbation is not essential due to the availability of rich training data.

Furthermore, when speech perturbation is not used, the results presented in Tables~\ref{tab:asru_results} and \ref{tab:seame_results} show that the proposed LAL can consistently improve the CS-ASR performance compared to the baseline system. Similar to other hyper-parameters such as learning rate and $\alpha$ in (\ref{eq:loss_asr}) and (\ref{eq:loss_all}) which require fine-tuning to achieve optimal performance, a fine-tuned $\beta$ results in significant improvements over the baseline model.

\begin{table*}[t]
\centering
\caption{Performance evaluation of the proposed method and state-of-the-art approaches on test data from the ASRU 2019 challenge and SEAME dataset in terms of English WER, Mandarin CER, and MER (\%). ``PT" denotes the model is pre-trained, ``$ft$" denotes fine-tuning, frame-level labels for ``LID'' are generated via forced alignment, ``FA-LB" and ``FA-UB” denote the forced alignment computed via fine-tuned Whisper-small (lower bound) and Whisper-large (upper bound) models, respectively}
\label{tab:sota_compare}
\renewcommand{\arraystretch}{1.1}
\setlength{\tabcolsep}{2.5mm}{
\begin{tabular}{c|c|c|ccc|ccc|ccc}
\toprule
\multirow{2}{*}{\textbf{Method}} & \multicolumn{1}{l|}{\textbf{\#Train.}} & \multirow{2}{*}{\textbf{PT}} & \multicolumn{3}{c|}{\textbf{ASRU test}} & \multicolumn{3}{c|}{\textbf{$\text{SEMAE test}_{\texttt{man}}$}} & \multicolumn{3}{c}{\textbf{$\text{SEAME test}_{\texttt{sge}}$}}  \\ 
 & \multicolumn{1}{l|}{\textbf{Params.}}&  & \textbf{Eng}$\downarrow$  & \textbf{Man}$\downarrow$ & \textbf{Mixed$\downarrow$} & \textbf{Eng} $\downarrow$ & \textbf{Man} $\downarrow$& \textbf{Mixed} $\downarrow$& \textbf{Eng} $\downarrow$& \textbf{Man} $\downarrow$& \textbf{Mixed} $\downarrow$\\ \midrule
\multicolumn{1}{l|}{Hybrid CTC/atten}            & 48.27 M  & \ding{55}    & 37.1  & 10.2  & 12.8  & 29.2  & 15.0  & 16.6  & 28.2 & 22.0 & 23.3 \\ 
\multicolumn{1}{l|}{+ LPB~\cite{liu23_icassp}}   & 79.90 M  & \ding{55}    & 35.3  & 9.22   & 11.8  & -     &  -    & 16.3  & -    & -    & 22.9 \\ 
\multicolumn{1}{l|}{+ LAL (ours)}                & 48.27 M  & \ding{55}    & 35.1  & 9.18   & 11.7  & 29.1  & 14.8  & 16.4  & 28.3 & 21.7 & 23.3 \\ \midrule
\multicolumn{1}{l|}{Bi-encoder~\cite{bi_encoder}}& 59.98 M  & \ding{55}    & 36.0  & 9.81   & 12.4  & 29.2  & 15.0  & 16.5  & 28.2 & 21.7 & 23.2 \\ \midrule 
\multicolumn{1}{l|}{Whisper-small~\cite{whisper}}& -        & \checkmark   & -     & -     & 24.9  & -     &  -    & 90.8  & -    &  -  & 69.7 \\ 
\multicolumn{1}{l|}{+ LoRA $ft$}                 & 3.24 M   & \checkmark   & -     & -     & 10.9  & -     &  -    & 16.0  & -    &  -  & 21.2 \\ 
\multicolumn{1}{l|}{+ $ft$}                      & 244.59 M & \checkmark   & -     & -     & 10.1  & -     &  -    & 14.3  & -    &  -  & 20.0 \\ 
\multicolumn{1}{l|}{+ $ft$ w/ prompt~\cite{yang2023adapting}}  & 244.59 M   & \checkmark   & -  & -    & 10.1     & -     &  -    & 15.1  & -     &  -  & 20.9\\ 
\multicolumn{1}{l|}{+ $ft$ w/ LID FA-LB}         & 244.59 M & \checkmark   & -     & -     & 9.91  & -     &  -    & 14.7  & -     &  -  & 19.9    \\ 
\multicolumn{1}{l|}{+ $ft$ w/ LID FA-UB}         & 244.59 M & \checkmark   & -     & -     & \textbf{9.61}  & -     &  -    & \textbf{14.3}  & -     &  -  & \textbf{19.3}    \\ 
\multicolumn{1}{l|}{+ $ft$ w/ LAL (ours)}        & 244.59 M & \checkmark   & -     & -     & 9.78  & -     &  -    & 14.5  & -   &  -  & 19.5    \\ \bottomrule

\end{tabular}}
\end{table*}

\subsection{Comparison with state-of-the-art methods}
We compare the performance of the proposed method against state-of-the-art approaches, including train-from-scratch and pre-trained models. These methods are evaluated on both model parameters and the CS-ASR performance is outlined in Table~\ref{tab:sota_compare}. For the ASRU dataset, the hybrid CTC/attention model incorporating LAL with language weights achieves the best MER of 11.7\% among the train-from-scratch approaches under consideration. While the bi-encoder and LPB methods exhibit higher performance compared to the baseline model, these approaches require an additional Conformer encoder and a Transformer decoder, respectively, resulting in a notable increase in model parameters. 

On the SEAME dataset, the hybrid CTC/attention model with the proposed LAL demonstrates a modest performance improvement compared to the vanilla model. Although the LPB method outperforms the proposed method, it requires a higher number of model parameters. In light of the ASRU dataset results, the above findings suggest that integrating the proposed LAL is a lightweight and efficient approach to attain high performance compared to other approaches.

The proposed method has also been validated through fine-tuning the Whisper model. The Whisper model initially suffers from low zero-shot performance on both datasets, with particularly high error rates of 90.8\% and 69.7\% observed on the SEAME test sets. However, models fine-tuned on the respective training sets demonstrate improved performance over the pre-trained Whisper, outperforming models trained from scratch. Fine-tuning with the proposed LAL leads to a higher performance compared to pure fine-tuning. This is consistent with the results observed in the hybrid CTC/attention model, highlighting the effectiveness of the proposed method.

Fine-tuning Whisper with either LID or LAL consistently results in improved performance compared to standard fine-tuning, indicating that incorporating auxiliary frame-level LID enhances CS-ASR performance. Furthermore, the Whisper model fine-tuned with the proposed LAL outperforms models that rely on auxiliary LID using language labels derived from FA-LB. This observation supports our statement in Section~\ref{sec:align_f_to_t} that the proposed LAL integrates the CS-ASR and LID into a mutual optimization process during training, leading to better performance than conventional LID approaches. However, the proposed LAL shows lower performance than jointly optimizing ASR and LID with FA-UB. This implies that the proposed LAL generates better pseudo labels compared to forced alignment computed by models with comparable performance levels, though these labels may still fall short of the accuracy provided by highly precise language labels, such as forced alignment computed by the fine-tuned Whisper-large model and gold-standard annotations.

\begin{table}[t]
\centering
\caption{Performance comparison of different types of linguistic hint used for prompting LLM via generative error correction by employing MER (\%) and utterance-level language identification accuracy (Acc \%) for English, Mandarin, and Code-switching. The N-best list is decoded from the hybrid CTC/attention model with LAL}
\label{tab:hints_results}
\renewcommand{\arraystretch}{1.1}
\setlength{\tabcolsep}{0.7mm}{
\begin{tabular}{c|cc|cc|cc}
\toprule
\textbf{LLM LoRA $ft$}  & \multicolumn{2}{c|}{\textbf{$\text{test}_{\texttt{man}}$}} & \multicolumn{2}{c|}{\textbf{$\text{test}_{\texttt{sge}}$}}     & \multicolumn{2}{c}{\textbf{ASRU $\text{test}$}}   \\ 
\textbf{Ling. hint type}& \textbf{MER} $\downarrow$  & \textbf{Acc.} $\uparrow$    & \textbf{MER} $\downarrow$ & \textbf{Acc.} $\uparrow$ & \textbf{MER} $\downarrow$  & \textbf{Acc.} $\uparrow$ \\ \midrule
\multicolumn{1}{l|}{no hint}         & 16.4          & -       & 23.3          & -    & 11.0                     & -    \\ 
\multicolumn{1}{l|}{LID output (LAL)}      & 17.0          & 80.3    & 24.4          & 84.3 & 11.0                     & 95.2 \\ 
\multicolumn{1}{l|}{1st hypo.}       & 16.5          & 93.3    & 23.1          & 92.5 & 11.0                     & 99.7 \\ 
\multicolumn{1}{l|}{hypos. vote}     & 16.6          & 93.2    & 23.0          & 93.0 & 11.0                     & 99.8 \\ 
\multicolumn{1}{l|}{LAL \& hypos. vote}& 16.6          & 92.9    & 23.1          & 92.5 & 11.0                     & 99.9 \\ 
\multicolumn{1}{l|}{groundtruth}     & \textbf{15.7} & -       & \textbf{22.0} & -    & \textbf{11.0}            & -    \\ \bottomrule
\end{tabular}}
\end{table}

\begin{table}[t]
\centering
\caption{Performance evaluation of prompting LLM with generative error correction after incorporating the linguistic hint by employing MER (\%). ``$ft$" denotes fine-tuning, ``$gt$" denotes the linguistic hint using ground-truth utterance-level language label, and ``$pred$" denotes the predicted linguistic hint}
\label{tab:llm_results}
\renewcommand{\arraystretch}{1.1}
\setlength{\tabcolsep}{1.8mm}{
\begin{tabular}{c|cc|c}
\toprule
\multirow{2}{*}{\textbf{Method}}      & \multicolumn{2}{c|}{\textbf{SEAME}}            & \textbf{ASRU} \\ 
                                      & \textbf{$\text{test}_{\texttt{man}}$}  & \textbf{$\text{test}_{\texttt{sge}}$}  & \textbf{test} \\ \midrule
\multicolumn{1}{l|}{Hybrid CTC/atten}               & 16.6 & 23.3      & 12.8  \\ 
\multicolumn{1}{l|}{+ LM $fusion$}             & 16.4 & 23.0      & 12.6  \\ \midrule
\multicolumn{1}{l|}{Hybrid CTC/atten w/ LAL}         & 16.4 & 23.3      & 11.7  \\ 
\multicolumn{1}{l|}{+ LM $fusion$}             & 16.4 & 23.1      & 11.9     \\ 
\multicolumn{1}{l|}{+ LLM LoRA $ft$ (no hint)}                & 16.4 & 23.3      & \textbf{11.0} \\ 
\multicolumn{1}{l|}{+ LLM LoRA $ft$ + ling. hint $(pred)$} & 16.5 & 23.1  & \textbf{11.0} \\ 
\multicolumn{1}{l|}{+ LLM LoRA $ft$ + ling. hint $(gt)$}   & \textbf{15.7} & \textbf{22.0}  & \textbf{11.0} \\ \bottomrule
\end{tabular}
}
\end{table}

\subsection{Incorporating LLM with linguistic hint}
To further enhance the CS-ASR performance, we integrate the LLM through prompting to perform generative error correction on the decoded N-best list. We compare the performance of the prompt in terms of MER and utterance-level LID accuracy before and after incorporating our proposed linguistic hints. The results on SEAME $\text{test}_{\texttt{man}}$, $\text{test}_{\texttt{man}}$, and the ASRU test set are presented in Table~\ref{tab:hints_results}. Here, the LID accuracy for the hypothesis within the N-best list is determined by summarizing the languages of tokens and that ``LID output" denotes the by-product of the proposed LAL.

Similar to~\cite{praveen23_interspeech, liu22e_interspeech}, it is not surprising that the LID accuracy achieved with the incorporation of textual information is higher than the by-product of the proposed LAL. This can be attributed to the fact that an ASR system models both acoustic and language characteristics, whereas an LID system generally focuses solely on acoustic information. Furthermore, a decrease in LID accuracy often results in a corresponding degradation in CS-ASR performance of the generated linguistic hints. The prompt incorporating the linguistic hint, which serves as the ground-truth language label, therefore exhibits significantly higher performance on the SEAME and ASRU datasets compared to other prompts. 

Notwithstanding the above, prompts with other linguistic hints achieve comparable performance to the prompt without the hint, albeit with a moderately lower overall performance in terms of MER. This observation suggests that linguistic hints may introduce a substantial bias into the generative error correction process. In particular, the utilization of a correct linguistic hint in a prompt enhances CS-ASR performance. Conversely, the misclassification of a linguistic hint results in an increased error rate. This implies that the proposed linguistic hint can potentially improve CS-ASR performance, particularly when an accurate language label is available.

The performance of CS-ASR systems with external language modeling is summarized in Table~\ref{tab:llm_results}. We observe that incorporating an LM via shallow fusion improves the CS-ASR performance. The use of LLM does not lead to performance improvement on the SEAME dataset unless accompanied by the ground-truth linguistic hint. In contrast, LLM can improve the performance on the ASRU dataset, achieving an MER of 11.0\% for all types of prompts. We analyzed the performance by considering the data distribution as shown in Table~\ref{tab:seame}. Since all utterances in the ASRU test set are code-switched, these predicted linguistic hints all exhibit high LID accuracy, and thus show comparable CS-ASR performance.

Since the LLM-based generative error correction method improves the CS-ASR performance over the baseline system on the ASRU data, results on the SEAME data imply that this performance mismatch may be due to differences between the two datasets.
As illustrated in Section~\ref{sec:datasets}, the SEAME dataset comprises spontaneous speech with frequent code-switching and approximately balanced English and Mandarin while the ASRU dataset consists of mostly read speech and Mandarin-dominant text. Since we employ a Chinese LLaMA model to perform general error correction, text data within the SEAME dataset can be significantly mismatched with the training data of Chinese LLaMA. This mismatch leads to a low performance of the LLM-based generative error correction method on the SEAME data. Moreover, incorporating the ground-truth linguist hint in the prompt results in significantly higher performance than other hints and the vanilla prompt. This suggests a possible solution for employing the LLM-based generative error correction method in a code-switching scenario, where a significant mismatch exists due to low resource or domain mismatch.

\begin{table}[t]
\centering
\caption{Performance comparison of prompting LLM with the proposed linguistic hint on $\text{test}_{\texttt{man}}$ set with (the original data) and without (removing or normalizing) interjections of the SEAME dataset by employing MER (\%) and utterance-level language identification accuracy (Acc \%). The N-best list is decoded from the hybrid CTC/attention model with LAL}
\label{tab:impact_interjection}
\renewcommand{\arraystretch}{1.1}
\setlength{\tabcolsep}{0.7mm}{
\begin{tabular}{c|cc|cc}
\toprule
\multirow{2}{*}{\textbf{Method}}  & \multicolumn{2}{c|}{\textbf{$\text{test}_{\texttt{man}}$}} & \multicolumn{2}{c}{$w/o$ $interjections$}    \\ 
   & \textbf{MER} $\downarrow$  & \textbf{Acc.} $\uparrow$    & \textbf{MER} $\downarrow$ & \textbf{Acc.} $\uparrow$ \\ \midrule
\multicolumn{1}{l|}{Hybrid CTC/atten w/ LAL}                                & 16.4 & -       & 13.6      & -    \\ \midrule
\multicolumn{1}{l|}{+ LLM LoRA $ft$ (no hint)}            & 16.4          & -       & 13.7      & -    \\ 
\multicolumn{1}{l|}{\hspace{1.5mm} w/ LID output (LAL)}   & 17.0          & 80.3    & 14.5      & 81.1 \\ 
\multicolumn{1}{l|}{\hspace{1.1mm} w/ 1st hypo.}          & 16.5    & \textbf{93.3} & 13.6      & \textbf{95.4} \\ 
\multicolumn{1}{l|}{\hspace{1.1mm} w/ hypos. vote}        & 16.8          & 93.2    & 13.7      & 95.3 \\ 
\multicolumn{1}{l|}{\hspace{1.1mm} w/ LAL \& hypos. vote} & 16.8          & 92.9    & 13.7      & 95.3 \\ 
\multicolumn{1}{l|}{\hspace{1.1mm} w/ groundtruth}        & \textbf{15.7} & -       & \textbf{13.4}  & -    \\ \bottomrule
\end{tabular}}
\end{table}

\begin{table}[t]
\centering
\caption{Performance evaluation of the proposed method on the ASRU data with different $\beta$ values in terms of utterance-level language identification accuracy (Acc. \%) and MER (\%)}
\label{tab:lid_acc}
\renewcommand{\arraystretch}{1.1}
\setlength{\tabcolsep}{3mm}{
\begin{tabular}{c|c|c}
\toprule
\textbf{$\beta$ for LAL} & \textbf{MER} $\downarrow$ & \textbf{Acc.} $\uparrow$\\ \midrule 
1.0& 12.1 & 95.5 \\ 
1.5& \textbf{11.9} & \textbf{95.7} \\ 
3.0& 12.0 & 94.1 \\ 
5.0& 12.2 & 93.4 \\ \bottomrule
\end{tabular}}
\end{table}

\subsection{Impact of interjections}
As described in Section~\ref{sec:datasets}, the SEAME dataset was collected from Singapore and Malaysia, where code-switching is more frequent than the ASRU dataset\textemdash interjections such as ``lah", ``lor", and ``ya" often occur in the former dataset. Annotating interjections is a challenging task even for individuals with bilingual expertise. Hence, interjections can introduce confusion during language modeling, leading to a degradation in the CS-ASR performance for LLMs that have not been trained on them. We therefore removed interjections from utterances in the SEAME training and $\text{test}_{\texttt{man}}$ without changing their semantic information. The CS-ASR performance and the LID accuracy of the linguistic hint prediction before and after removing interjections when incorporating LLM are shown in Table~\ref{tab:impact_interjection}. In addition, the first row shows the MER of the hybrid CTC/attention model with LAL. This is computed by comparing the decoded hypotheses and ground-truth transcriptions after removing interjections within them.

The results indicate that removing interjections improves the CS-ASR performance significantly due to the reduction of language confusion. However, incorporating the LLM does not benefit the CS-ASR performance except for further use of the linguistic hint with ground-truth language labels. This is consistent with results presented in Tables~\ref{tab:hints_results} and \ref{tab:llm_results}, suggesting that a correct linguistic hint can lead to performance improvement.

\subsection{Language identification performance}
As a key component of the proposed LAL method, the LID branch often benefits from the ASR module. However, experiments in existing works have shown that LID, when employed as an auxiliary task, may not necessarily improve the ASR performance~\cite{zeng19_interspeech,liu23_icassp,10095326} when the LID-related parameter (i.e., $\beta$) increases.

Our primary objective is to improve the CS-ASR performance of the model with the assistance of language information instead of achieving high LID performance. Although the proposed LAL generally improves the CS-ASR performance, tuning an appropriate $\beta$ for the LAL loss term in (\ref{eq:loss_all}) is crucial for optimal performance improvement. Therefore, we present the LID performance of models trained with various $\beta$ values in Table~\ref{tab:lid_acc}.

As shown in the table, the model achieves the lowest LID performance when $\beta=5.0$. This suggests that using a higher $\beta$ value may not necessarily achieve the optimal LID performance. The highest LID performance is achieved with $\beta=1.5$. This indicates that the model optimized with an appropriate $\beta$ value can achieve high performance for both CS-ASR and LID tasks. In addition, this implies that higher ASR performance can improve the LID branch.

\section{Discussion}
\label{sec:discuss}
\subsection{Frame-to-token alignment}
\textbf{CTC.} In the hybrid CTC/attention ASR model, the frame-to-token alignment can be computed from both CTC and the cross-attention process within the ASR decoder. However, CTC predictions contain \textit{$<$blank$>$} token that results in peaky behavior~\cite{zeyer2021does, chen23c_interspeech}. Since the \textit{$<$blank$>$} token lacks a language attribute and cannot be converted into a language label, this peaky behavior leads to fewer language labels in the pseudo label sequences compared to those derived through the ASR decoder. Therefore, we use the frame-to-token alignment computed from the cross-attention weight matrix within the ASR decoder but not from CTC to generate pseudo-language labels.

\textbf{Forced alignment.} We can also use additional processing (e.g., forced alignment) to retrieve token-level timestamps and achieve frame-to-token alignment. However, state-of-the-art multilingual ASR models, such as Whisper, require fine-tuning to achieve satisfactory performance. As presented in Table~\ref{tab:sota_compare}, the pre-trained Whisper-small model achieves MERs of 24.9\%, 90.8\%, and 69.7\% on the ASRU and SEAME test sets, respectively. As a result, applying forced alignment to code-switching speech using pre-trained multilingual ASR models is undesirable.

Results presented in Table~\ref{tab:sota_compare} show that the proposed LAL method outperforms applying LID with forced-alignment language labels computed by the fine-tuned Whisper-small model (FA-LB). Moreover, forced alignment incurs a substantial computational burden, as it must be applied to the entire training dataset using fine-tuned Whisper models. While LID with FA-UB exhibits higher performance than the proposed LAL, it results in an even greater computational cost than FA-LB due to its reliance on the Whisper-large model. This contradicts the motivation of our work, which is to enrich the CS-ASR model with language information without the need for an additional annotating process. 

\begin{table}[t]
\centering
\caption{Comparison of the original ASR with LAL outputs, the LM late fusion outputs, the LLM-generated outputs with ground-truth linguistic hint, and the ground-truth transcripts. The first four examples are from the SEAME dataset, and the last one is from the ASRU dataset}
\label{tab:where_error_happen}
\setlength{\tabcolsep}{0.9mm}{
\begin{tabular}{l|l}
\toprule
\multicolumn{1}{c|}{\textbf{Method}}  & \multicolumn{1}{c}{\textbf{Output}}               \\ \midrule
\multirow{5}{*}{ASR w/ LAL}           & 唉 呀                                               \\
                                      & ah yeah close already                             \\
                                      & the yeah what happen to him ah   \\
                                      & but 你 先 熬 一 年 先 啦                                 \\ 
                                      & 在 雨 田 机 场 体 验 了 一 下 shower room \\ \midrule
\multirow{5}{*}{+ LM $late fusion$}   & 唉 呀                            \\
                                      & ah yah close already    \\
                                      & the yeah what happen to him ah \\
                                      & but 你 先 熬 一 年 先 啦                                   \\ 
                                      & 在 雨 田 机 场 体 验 了 一 下 shower room \\ \midrule
\multirow{5}{*}{+ LLM LoRA $ft$}      & ah yah                                                   \\
                                      & ah you are close already                                 \\
                                      & the yeah what happened to him ah                         \\
                                      & but 你 现 在 熬 一 年 先 啦                                 \\ 
                                      & 在 羽 田 机 场 体 验 了 一 下 shower room  \\ \midrule
\multirow{5}{*}{Groundtruth}          & ah yeah                                                   \\
                                      & ah yeah close with me                                     \\
                                      & the yeah what happened to him hah \\
                                      & but 你 先 熬 一 年 先 啦          \\ 
                                      & 在 羽 田 机 场 体 验 了 一 下 shower room \\ \bottomrule
\end{tabular}}
\end{table}

\subsection{Where the errors happen}
The errors have been analyzed to gain insight into the factors affecting ASR performance. Compared to the deletion and insertion rates, a significantly higher substitution rate is observed from the aforementioned results. The high substitution rate imposes a challenge in language modeling due to a larger vocabulary and language confusion arising from a code-switching scenario. Given that code-switched text may be more readily achieved than speech, developing a robust code-switching language model is desirable to address language confusion effectively.

In addition to the model performance in CS-ASR, the performance in terms of token-level language identification is worth highlighting. Compared to the CS-ASR model trained on the SEAME dataset, the model trained on the ASRU dataset can generally identify the token-level language change points. This underpins that the SEAME dataset is more challenging than the ASRU dataset since the two languages are less discriminative in the SEAME dataset. 

To gain further insights into the CS-ASR performance, we compared the decoded outputs of the ASR model incorporating LAL and its counterparts enhanced by LM and LLM. The results are juxtaposed with the ground-truth transcription in Table~\ref{tab:where_error_happen}. One factor that contributes to errors in the CS-ASR task is language confusion. For instance, the expression ``ah yeah" shares both the pronunciation and semantic characteristics with its Chinese equivalent ``唉 呀". Another notable factor that contributes to errors is the liaison, where two Mandarin characters or English words can erroneously be classified as a single entity. The liaison effect can be particularly pronounced in a spontaneous code-switching speech signal. 

The fine-tuned LLM demonstrated effectiveness in correcting grammatical errors for SEAME samples. Words such as ``happened" are consequently adjusted to the correct tense. However, the model is less adept at accommodating colloquial expressions. Therefore, the third and fourth cases illustrated in Table~\ref{tab:where_error_happen} have been modified to a more formal expression, leading to a higher mixed error rate. For the ASRU example, the fine-tuned LLM successfully corrects the incorrectly predicted entity name, likely due to its contextual understanding and world knowledge acquired during large-scale pre-training.
\begin{figure}[t]
  \centering
  \includegraphics[width=\linewidth]{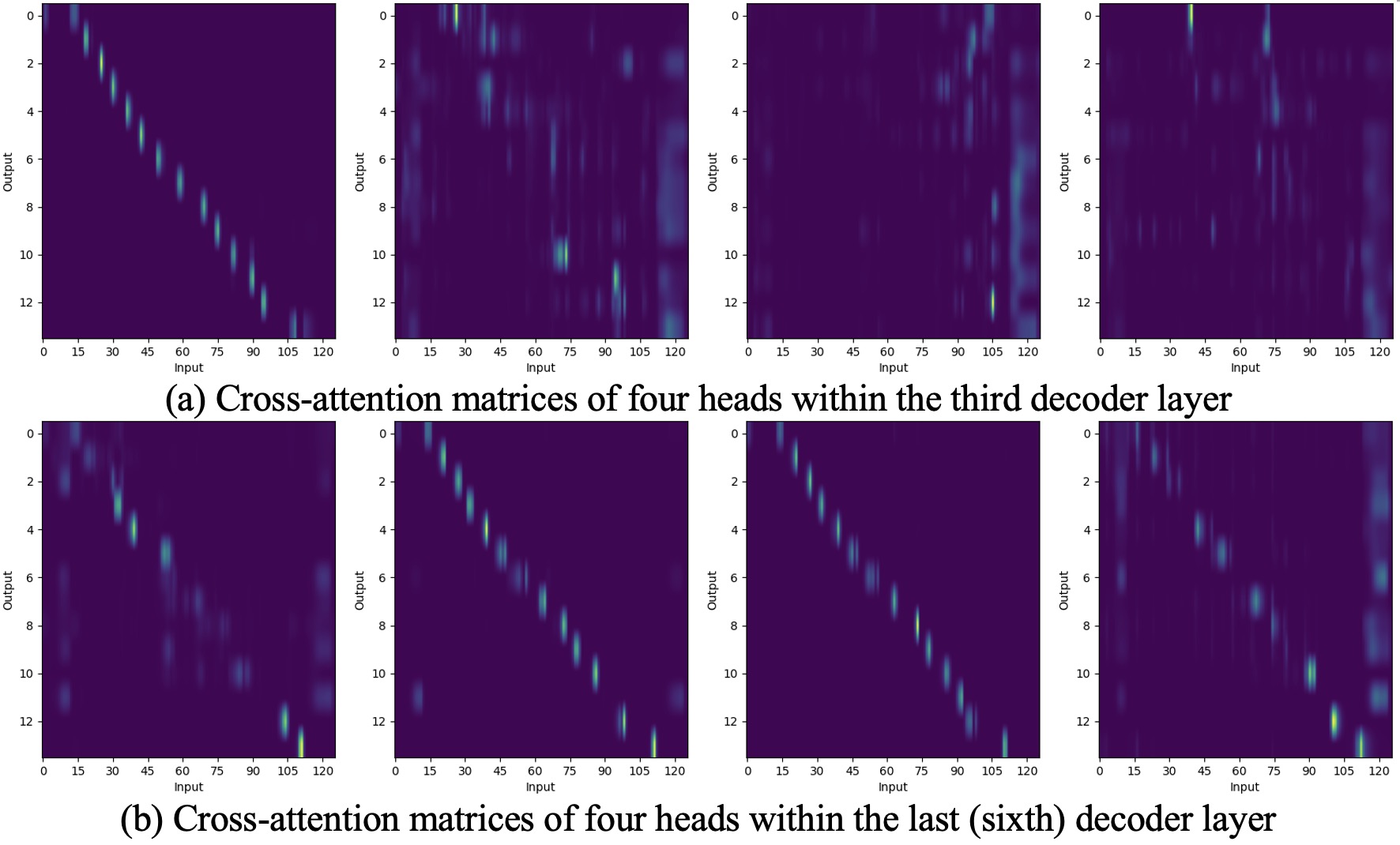}
  \caption{The visualization of multi-head cross-attention matrices within the ASR decoder layers, where the x- and y-axis denote speech frames (encoder output) and tokens, respectively.}
  \label{fig:atten_matrix}
\end{figure}

\subsection{Attention matrices}
We visualize the multi-head attention matrices within various ASR decoder layers corresponding to a speech sample as shown in Fig. 6, where the abscissa and the ordinate axes denote speech frames (encoder output) and tokens, respectively. The model is the best-performing CS-ASR model with the proposed LAL on the ASRU data.

Existing works have shown that shallow layers within a Transformer-based model focus on low-level speech information such as phonetics and language, while deep layers often capture high-level information such as semantic and sequential characteristics~\cite{ssl_review, wavlm}. This is consistent with the visualization, motivating us to employ the last decoder layer. Therefore, the average of cross-attention matrices within the last decoder layer is used to compute the frame-level pseudo-language labels.

Specifically, compared to the attention matrices in the third decoder layer, those in the last decoder layer capture sequential information more effectively and display a monotonic mapping between frames and tokens. However, Fig. 6(b) shows that the monotonic mapping can be blurred due to the difficulty in learning the attention pattern of each attention head during training. To mitigate this challenge, the averaged frame-to-token alignment across all attention heads is used instead of relying on a single attention head.

\subsection{LAL in Speech LLMs}
The proposed method holds potential for integration into both the pre-training and fine-tuning stages of multi-modal LLMs, such as speech LLMs, that support speech input, particularly for code-switching scenarios.

The speech encoder in speech LLMs is typically derived from the encoder of a large-scale pre-trained ASR model~\cite{Qwen-Audio, he2024meralion, whisper}. Incorporating the proposed LAL into the speech encoder or tokenizer~\cite{whisper, zhang2024speechtokenizer} during pre-training on code-switching data may effectively enrich the model with code-switching capability. This integration can help the model better capture language boundaries and transitions, which are particularly prevalent in Southeast Asian multilingual contexts. Moreover, while large-scale ASR models already achieve strong recognition performance, their encoders produce rich speech representations that can support more accurate LID decisions for LAL. These LID decisions can then be integrated into the prompt, providing the LLM with explicit language guidance during text generation.

However, the proposed LAL may not be well-suited for general-purpose speech LLMs in which code-switching data rarely exists in the training corpus. In such settings, the supervision provided by LAL would largely resemble an auxiliary utterance-level language identification task, offering limited benefit. As a result, incorporating LAL into speech LLMs may not yield meaningful improvements in these scenarios.

Furthermore, the LLM module within speech LLMs can internally perform generative error correction before generating the final ASR output. This correction is achieved through the reasoning capabilities of LLMs, applied to either the N-best hypotheses or a single hypothesis during inference. Recent LLMs have demonstrated stronger multilingual capabilities in both understanding and reasoning compared to earlier models such as LLaMA-2 used in this work~\cite{team2024gemma, llama3}. These advancements can enhance internal error correction in code-switching ASR, particularly when guided by the proposed linguistic hint.

\section{Conclusion}
\label{sec:conclusion}
We proposed to align speech to languages to enhance the CS-ASR performance in both ASR and LLM-based processing. For the ASR model, we introduced a language alignment loss~(LAL) to enrich the model with language information. Models equipped with the proposed LAL consistently achieve higher CS-ASR performance than the vanilla configuration, with only a negligible increase in model parameters during training. In addition, the proposed LAL outperforms frame-level LID using language labels obtained through forced alignment. These demonstrated the effectiveness of the proposed LAL. After incorporating language weights into LAL to address language imbalance in the ASRU data during training, the proposed method obtained further performance improvement. Beyond ASR, we proposed leveraging a linguistic hint, which is derived from LAL outputs and decoded hypotheses, to guide the prompting and enhance the LLM-based generative error correction. Experimental results indicate that an accurate linguistic hint can significantly improve CS-ASR performance in scenarios involving both monolingual and code-switching utterances. Finally, the errors within the hypotheses are analyzed. The LLM fine-tuned on the SEAME data has shown effectiveness in correcting grammatical errors, which, in contrast, leads to lower ASR performance for spontaneous and colloquial speech. 

\normalem
\bibliographystyle{IEEEtran}
\bibliography{refs}
\end{CJK}
\end{document}